\documentstyle[version2,preprint,prl,aps]{revtex}
\begin{document}
\draft
\rightline{WIS-92/101/DEC-PH\quad}
\begin{title}
Levinson's Theorem for Dirac Equation
\end{title}
\author{Nathan Poliatzky\cite{Email}}
\begin{instit}
Department of Physics, The Weizmann Institute of Science,
Rehovot, Israel
\end{instit}
\begin{abstract}
Levinson's theorem for the Dirac equation is known in the form of a
sum of positive and negative energy phase shifts at zero momentum
related to the total number of bound states. In this letter we prove
a stronger version of Levinson's theorem valid for positive and
negative energy phase shifts separately. The surprising result is,
that in general the phase shifts for each sign of the energy do not
give the number of bound states with the same sign of the energy (in
units of $\pi$), but instead, are related to the number of bound
states of a certain Schr\"odinger equation, which coincides with the
Dirac equation at zero momentum.
\end{abstract}
\pacs{PACS numbers: 0365}

\narrowtext

Consider the reduced radial Schr\"odinger equation
\begin{equation}
u''_{kl}-\bigg\lbrack\frac{l(l+1)}{r^2}+2mV-k^2\bigg\rbrack u_{kl}=0\,,
\label{schreq}
\end{equation}
for a scattering state characterized by the reduced radial wave
function $u_{kl}(r)$ subject to the boundary conditions
\begin{equation}
u_{kl}(0)=0
\end{equation}
at the origin and
\begin{equation}
u_{kl}(r)\rightarrow2\sin\left(kr-\frac{\pi l}{2}+\eta_l(k)\right)
\end{equation}
at infinity ($r\rightarrow\infty$), where $\eta_l(k)$ is the phase
shift. Here we assume that the potential $V(r)$ is less singular
at the origin than $1/r^2$ and that it vanishes at infinity faster
than $1/r$. Levinson's theorem \cite{Levinson} asserts that
\begin{equation}
\eta_l(0)=n_l\pi\,,
\label{Lev}\end{equation}
and thus establishes a connection between the scattering phase shift
$\eta_l(0)$ at threshold (zero momentum) and for a given
angular momentum $l$ to the number of bound states $n_l$ of the
Schr\"odinger equation (\ref{schreq}). If the Schr\"odinger equation
has a zero-energy solution which vanishes at the origin and is finite
at infinity and yet not normalizable (it is called a half-bound
state or zero-energy resonance and is possible only if $l=0$) then,
as was first shown by R. Newton \cite{Newtona}, Levinson's theorem
is modified to read
\begin{equation}
\eta_0(0)=\left(n_0+\frac{1}{2}\right)\pi\,.
\label{LevNewt}\end{equation}
Levinson's theorem, which turned out to be fairly general
\cite{Jauch,polscatt}, is one of the most interesting results
in quantum theory. It has many potential applications and has been
applied recently in atomic physics \cite{Spruchprl,Spruchpr},
in quantum field theories \cite{Blank,Niemi}, and in solid state
physics (where it is known in a modified form under the name Friedel's
sum rule \cite{Friedel}).

Consider now the reduced radial Dirac equation
\begin{eqnarray}
&&u'_{1\epsilon\kappa}+\frac{\kappa}{r} u_{1\epsilon\kappa}-
\left(\epsilon+m-V\right)u_{2\epsilon\kappa}=0
\nonumber\\&&\label{raddir}\\
&&u'_{2\epsilon\kappa}-\frac{\kappa}{r}\, u_{2\epsilon\kappa}+
\left(\epsilon-m-V\right)u_{1\epsilon\kappa}=0\,,
\nonumber\end{eqnarray}
where $\epsilon$ is the energy, $V(r)$ is the time
component of a vector potential and $\kappa=\pm1,\pm2,\ldots\,$.
The quantum number $\kappa$ is the standard parametrization of the
total angular momentum $j=\vert\kappa\vert-1/2$ and of the relative
orientation between the spin and the orbital angular momentum. The
appropriate boundary conditions at the origin and infinity are
\begin{eqnarray}
&&u_{1\epsilon\kappa}\left(0\right)=0\,,\nonumber\\&&\label{bcondo}\\
&&u_{1\epsilon\kappa}\left(r\right)\rightarrow\sqrt{\frac{\epsilon+m}
{2\epsilon}}2\sin\left(kr-\frac{\pi l}{2}+\eta_{\epsilon\kappa}
\left(k\right)\right)\,,
\nonumber\end{eqnarray}
\begin{eqnarray}
&&u_{2\epsilon\kappa}\left(0\right)=0\nonumber\\&&\label{bcondt}\\
&&u_{2\epsilon\kappa}\left(r\right)\rightarrow\sqrt{\frac{\epsilon+m}
{2\epsilon}}\frac{2\kappa_0k}{\epsilon+m}\sin\left(kr-\frac
{\pi\overline{l}}{2}+\eta_{\epsilon\kappa}\left(k\right)\right),
\nonumber\end{eqnarray}
where $k=\pm\sqrt{\epsilon^2-m^2}$, $\kappa_0=\kappa/\vert\kappa
\vert$, $l=\vert\kappa\vert-(1-\kappa_0)/2$, $\overline l=l-\kappa_0$,
and $\eta_{\epsilon\kappa}\left(k\right)$ is the phase shift. To
ensure the consistency of (\ref{bcondo}-\ref{bcondt}) with
(\ref{raddir}) we assume that $V(r)$ behaves like or less singularly
than $1/r$ at the origin and that it vanishes at infinity faster than
$1/r$. The first correct statement of Levinson's theorem for Dirac
particles was given by Ma and Ni \cite{MaNi}
\begin{equation}
\eta_{m\kappa}\left(0\right)+\eta_{-m\kappa}\left(0\right)=\left(
N_\kappa^++N_\kappa^-\right)\pi\,,
\label{Levd}\end{equation}
which is valid whenever there is no threshold resonance and
\begin{eqnarray}
\eta_{m\kappa}\left(0\right)&+&\eta_{-m\kappa}\left(0\right)=\left(
N_\kappa^++N_\kappa^-\right)\pi\nonumber\\& &\label{Levmodd}\\
&+&(-1)^l\frac{\pi}{2}\left(\sin^2\eta_{m\kappa}\left(0\right)+\sin^2
\eta_{-m\kappa}\left(0\right)\right)\,,
\nonumber\end{eqnarray}
which is valid for the case with a threshold resonance (which can
appear only in the case $\kappa=\pm1$, see \cite{polscatt}). Here
$\pm m$ is the threshold energy of the Dirac particle, $N_\kappa^+$
is the number of positive and $N_\kappa^-$ the number of negative
energy bound states of the Dirac equation (\ref{raddir}) and
$\eta_{\pm m\kappa}(0)$ are the phase shifts at
threshold. Prior to the work of Ma and Ni claims were published
\cite{Barth,Ni} stating that Levinson's theorem
is valid for positive and negative energies separately and
in the same sense as in the nonrelativistic case, i.e.
$\eta_{\pm m\kappa}(0)=N_\kappa^\pm\pi$, but later such claims were
found incorrect \cite{MaNi}. However, we shall prove in this letter
that in a modified sense these claims are correct and that
\begin{eqnarray}
\eta_{m\kappa}(0)&=&\left\{
\begin{array}{l}
n_l^+\pi\,,\hskip50pt l=0,1,\ldots\\ \\
\left(n_0^++\frac{1}{2}\right)\pi\,,\hskip20pt l=0\,,
\end{array}\right.\label{Levwdsolp}\\& &\nonumber\\
& &\nonumber\\
\eta_{-m\kappa}(0)&=&\left\{
\begin{array}{l}
n_{\overline l}^-\pi\,,\hskip50pt\overline l=0,1,\ldots\\ \\
\left(n_0^-+\frac{1}{2}\right)\pi\,,\hskip20pt\overline l=0\,,
\end{array}\right.\label{Levwdsolm}
\end{eqnarray}
where $n_l^+$ and $n_{\overline l}^-$ are the numbers of bound state
solutions of certain radial Schr\"odinger equations with the angular
momenta $l$ and $\overline l=l-\kappa/\vert\kappa\vert$. In
(\ref{Levwdsolp}) and (\ref{Levwdsolm}) the first case refers to a
situation without a threshold resonance and the second case to a
situation with a threshold resonance. Equations (\ref{Levwdsolp})
and (\ref{Levwdsolm}) constitute the stronger version of
Levinson's theorem for Dirac particles. Notice that from (\ref{Levd})
and (\ref{Levwdsolp}), (\ref{Levwdsolm}) it follows that
\begin{equation}
N_\kappa^++N_\kappa^-=n_l^++n_{\overline l}^-\,,
\label{Nnpm}\end{equation}
whereas in general $N_\kappa^+\neq n_l^+$ and
$N_\kappa^-\neq n_{\overline l}^-$. Below it will be shown that
(\ref{Nnpm}) is actually a general statement (independent of
Levinson's theorem).

The basic observation, which we need in order to derive
(\ref{Levwdsolp}) and (\ref{Levwdsolm}), is the fact that one can
write the Dirac equation (\ref{raddir}) as a set of Schr\"odinger-like
equations. In fact, eliminating $u_{2\epsilon\kappa}$ from
(\ref{raddir}), we obtain
\begin{eqnarray}
u_{1\epsilon\kappa}''-\bigg\lbrack&&\frac{\kappa\left(\kappa+1
\right)}{r^2}-\frac{\kappa}{r}\frac{V'}{\epsilon+m-V}-V^2
\nonumber\\&&\label{qdireqo}\\
&&+2\epsilon V-k^2\bigg\rbrack u_{1\epsilon\kappa}+\frac{V'}
{\epsilon+m-V}\,u_{1\epsilon\kappa}'=0\,.
\nonumber\end{eqnarray}
Eliminating $u_{1\epsilon\kappa}$ from (\ref{raddir}) leads to
\begin{eqnarray}
u_{2\epsilon\kappa}''-\bigg\lbrack&&\frac{\kappa\left(\kappa-1
\right)}{r^2}+\frac{\kappa}{r}\frac{V'}{\epsilon-m-V}-V^2
\nonumber\\&&\label{qdireqt}\\
&&+2\epsilon V-k^2\bigg\rbrack u_{2\epsilon\kappa}+\frac{V'}
{\epsilon-m-V}\,u_{2\epsilon\kappa}'=0\,.
\nonumber\end{eqnarray}
Equations (\ref{qdireqo}) and (\ref{qdireqt}) are equivalent to the
Dirac equation (\ref{raddir}) provided the boundary conditions
(\ref{bcondo}) and (\ref{bcondt}) are imposed. Actually it is not
necessary to solve (\ref{qdireqo}) and (\ref{qdireqt}) simultaneously.
For instance, if (\ref{qdireqo}) is solved for $u_{1\epsilon\kappa}$
one gets $u_{2\epsilon\kappa}$ through the first of equations
(\ref{raddir}). In order to get rid of the last term in
(\ref{qdireqo}) and (\ref{qdireqt}) we introduce the new wave
functions $w_{1\epsilon\kappa}$, $w_{2\epsilon\kappa}$ defined
through
\begin{equation}
u_{1\epsilon\kappa}\left(r\right)=\sqrt{\frac{\epsilon+m-V\left(r
\right)}{2\epsilon}}w_{1\epsilon\kappa}\left(r\right)\,,
\label{wodir}\end{equation}
\begin{equation}
u_{2\epsilon\kappa}\left(r\right)=\frac{\epsilon k\kappa}{\vert
\epsilon\vert\vert k\vert\vert\kappa\vert}\sqrt{\frac{\epsilon-m-V
\left(r\right)}{2\epsilon}}w_{2\epsilon\kappa}\left(r\right)\,.
\label{wtdir}\end{equation}
Putting these in (\ref{qdireqo}) and (\ref{qdireqt}), we obtain
\begin{eqnarray}
w_{1\epsilon\kappa}''-\bigg\lbrack&&\frac{l(l+1)}{r^2}-\frac{\kappa}
{r}\frac{V'}{\epsilon+m-V}+\frac{1}{2}\frac{V''}{\epsilon+m-V}
\nonumber\\
&&+\frac{3}{4}\left(\frac{V'}{\epsilon+m-V}\right)^2-V^2
\label{wdireqo}\\
&&\hskip65pt+2\epsilon V-k^2\bigg\rbrack w_{1\epsilon\kappa}=0\,,
\nonumber
\end{eqnarray}
\begin{eqnarray}
w_{2\epsilon\kappa}''-\bigg\lbrack&&\frac{\overline l\left(
\overline l+1\right)}{r^2}+\frac{\kappa}{r}\frac{V'}{\epsilon-m-V}+
\frac{1}{2}\frac{V''}{\epsilon-m-V}\nonumber\\
&&+\frac{3}{4}\left(\frac{V'}{\epsilon-m-V}\right)^2-V^2
\label{wdireqt}\\
&&\hskip65pt+2\epsilon V-k^2\bigg\rbrack w_{2\epsilon\kappa}=0\,,
\nonumber\end{eqnarray}
where we used $\kappa\left(\kappa+1\right)=l\left(l+1\right)$ and
$\kappa\left(\kappa-1\right)=\overline l\left(\overline l+1\right)$
which are readily obtained from $l=\vert\kappa\vert-\left(1-\kappa_0
\right)/2$ and $\overline l=l-\kappa_0$, $\kappa_0=\kappa/\vert\kappa
\vert$. Equations (\ref{wdireqo}) and (\ref{wdireqt}) are of the
Schr\"odinger-type except that the potential depends on the energy.
Solving (\ref{wdireqo}) and (\ref{wdireqt}) is equivalent to solving
the original Dirac equation (\ref{raddir}) provided the boundary
conditions at the origin and infinity
\begin{eqnarray}
&&w_{1\epsilon\kappa}\left(0\right)=0\,,\nonumber\\&&\label{wbcondo}\\
&&\quad w_{1\epsilon\kappa}\left(r\right)\rightarrow2\sin\left(kr-
\frac{\pi l}{2}+\eta_{\epsilon\kappa}\left(k\right)\right)\,,
\nonumber\end{eqnarray}
\begin{eqnarray}
&&w_{2\epsilon\kappa}\left(0\right)=0\,,\nonumber\\&&\label{wbcondt}\\
&&\quad w_{2\epsilon\kappa}\left(r\right)\rightarrow2\sin\left(kr-
\frac{\pi\overline l}{2}+\eta_{\epsilon\kappa}\left(k\right)\right)
\nonumber\end{eqnarray}
are taken into account. Equations (\ref{wdireqo}) and (\ref{wdireqt})
are useful because they are not coupled and hence the phase shift
$\eta_{\epsilon\kappa}\left(k\right)$ can be computed using
any one of them, without reference to the other and for each of the
positive and negative energies $\epsilon$ separately. Yet, we
cannot apply Levinson's theorem to (\ref{wdireqo}) or (\ref{wdireqt})
directly since the potential in these equations depends on the
energy, and one can show that in this case the theorem is not valid
\cite{polscatt}. However, consider the following equations
\begin{eqnarray}
w_{kl}^+{''}-\bigg\lbrack&&\frac{l\left(l+1\right)}{r^2}-\frac{\kappa}
{r}\frac{V'}{2m-V}+\frac{1}{2}\frac{V''}{2m-V}\nonumber\\
&&+\frac{3}{4}\left(\frac{V'}{2m-V}\right)^2-V^2\label{wdireqp}\\
&&\hskip65pt+2mV-k^2\bigg\rbrack w_{kl}^+=0\,,
\nonumber\end{eqnarray}
\begin{eqnarray}
w_{k\overline l}^-{''}-\bigg\lbrack&&\frac{\overline l\left(
\overline l+1\right)}{r^2}-\frac{\kappa}{r}\frac{V'}{2m+V}-\frac{1}{2}
\frac{V''}{2m+V}\nonumber\\
&&+\frac{3}{4}\left(\frac{V'}{2m+V}\right)^2-V^2\label{wdireqm}\\
&&\hskip65pt-2mV-k^2\bigg\rbrack w_{k\overline l}^-=0\,,
\nonumber\end{eqnarray}
which are subject to the boundary conditions
\begin{eqnarray}
&&w_{kl}^+\left(0\right)=0\,,\nonumber\\&&\label{wbcondp}\\
&&\quad w_{kl}^+\left(r\right)\rightarrow2\sin\left(kr-\frac{\pi l}{2}
+\eta_l^+\left(k\right)\right)\,,
\nonumber\end{eqnarray}
\begin{eqnarray}
&&w_{k\overline l}^-\left(0\right)=0\,,\nonumber\\&&\label{wbcondm}\\
&&\quad w_{k\overline l}^-\left(r\right)\rightarrow2\sin\left(kr-
\frac{\pi\overline l}{2}+\eta_{\overline l}^-\left(k\right)\right)\,.
\nonumber\end{eqnarray}
At threshold ($k=0$) (\ref{wdireqp}) and (\ref{wbcondp}) coincide with
(\ref{wdireqo}) and (\ref{wbcondo}) for $\epsilon=m$, and similarly
(\ref{wdireqm}) and (\ref{wbcondm}) coincide with (\ref{wdireqt}) and
(\ref{wbcondt}) for $\epsilon=-m$. Moreover, both sets of equations
and boundary conditions are analytical near the threshold. Therefore
\begin{equation}
\eta_l^+\left(0\right)=\eta_{m\kappa}\left(0\right)
\label{etaplus}\end{equation}
and
\begin{equation}
\eta_{\overline l}^-\left(0\right)=\eta_{-m,\kappa}\left(0\right)\,.
\label{etaminus}\end{equation}
Equations (\ref{wdireqp}) and (\ref{wdireqm}) are just usual
Schr\"odinger equations, linear in the energy $k^2$, so that we can
apply Levinson's theorem (\ref{Lev}) and (\ref{LevNewt}) and obtain
the desired equations (\ref{Levwdsolp}) and (\ref{Levwdsolm}), where
$n_l^+$ is the number of bound state solutions ($k^2<0$) of
(\ref{wdireqp}) and $n_{\overline l}^-$ is the number of bound state
solutions of (\ref{wdireqm}). Actually one would expect an ambiguity
in (\ref{etaplus}) and in (\ref{etaminus}), each in terms of an
additive integer multiple of $\pi$. However, it is easy to see that
both integers (say $n_1$ and $n_2$) must be zero. This follows from
the fact that the simultaneous change  of $m$ to $-m$ and $\kappa$ to
$-\kappa$ is a symmetry operation, which implies $n_1=n_2$.
Equations (\ref{Levd}) and (\ref{Nnpm}), on the other hand, imply
$n_1=-n_2$, and hence both integers are zero. To prove that (\ref{Nnpm})
is valid generally (independently of Levinson's theorem) we make the
following observation. Let us multiply the potential $V$ in
(\ref{wdireqo}-\ref{wdireqt}) and in (\ref{wdireqp}-\ref{wdireqm})
by a coupling constant $g$. For $g=0$ there are no bound states and
(\ref{Nnpm}) is (trivially) valid. We now change $g$ continuously
from $0$ to $1$ and show that (\ref{Nnpm}) remains valid. Assume that
(\ref{Nnpm}) is valid for some value of $g$. Then, if $g$ is
increased, at some point a (half-) bound state will either appear or
disappear. This happens simultaneously for (\ref{wdireqo}) and
(\ref{wdireqp}), or/and for (\ref{wdireqt}) and (\ref{wdireqm}) and in
the same direction, for the corresponding equations are equal at and
analytic near $k=0$. Hence the process of bound states
entering or leaving through the point $k=0$ does not alter the
validity of (\ref{Nnpm}). But this is not all. Parallel to this
process there is a motion of bound states of the Dirac equation
(\ref{wdireqo}-\ref{wdireqt}) through the point $\epsilon=0$
(equivalently $k^2=-m^2$). This process, however, does not change the
total number of bound states since, here, if a positive energy bound
state disappeares, simultaneously a negative energy bound state
appears or vice versa. These two processes are all what happens to the
number of bound states, if we drive $g$ from $0$ to $1$. Consequently,
we arive safely at $g=1$ with (\ref{Nnpm}) still being valid.

It is an interesting fact to notice that (\ref{wdireqp}) and
(\ref{wdireqm}) do not correspond to the usual expansion based on the
Foldy-Wouthuysen scheme (see \cite{F-W}). More details and an
application of Levinson's theorem to a nonperturbative approach to
quantum electrodynamics will be given in \cite{polscatt}.

\acknowledgments

The author wishes to thank Y. Eisenberg and V. Nabutovsky for valuable
discussions and suggestions. Thanks are also due to R. G. Newton for
clarifying remarks.


\begin{references}
\bibitem[*]{Email}{Present address: Theoretische Physik,
ETH-H\"onggerberg, CH-8093 Z\"urich, Switzerland. E-mail:
poli@itp.ethz.ch}
\bibitem{Levinson}{N. Levinson, Kgl. Danske Videnskab. Selskab,
Mat.-fys. Medd., {\bf 25}(9) (1949).}
\bibitem{Newtona}{R. G. Newton, J. Math. Phys. {\bf 1}, 319 (1960).}
\bibitem{Jauch}{ J. M. Jauch, Helv. Phys. Acta, {\bf30}, 143 (1957).}
\bibitem{polscatt}{N. Poliatzky, Normalization of scattering states,
scattering phase shifts and Levinson's theorem. To appear in
Helv. Phys. Acta. (1993).}
\bibitem{Spruchprl}{Z. R. Iwinski, Leonard Rosenberg, and Larry
Spruch, Phys. Rev. Lett., 1602 (1985).}
\bibitem{Spruchpr}{Z. R. Iwinski, Leonard Rosenberg, and Larry
Spruch, Phys. Rev. A {\bf33}, 946 (1986).}
\bibitem{Blank}{R. Blankenbecler and D. Boyanovsky, Physica 18D, 367
(1986).}
\bibitem{Niemi}{A. J. Niemi and G. W. Semenoff, Phys. Rev. D {\bf32},
471 (1985).}
\bibitem{Friedel}{J. Friedel, Nuovo Cim. Suppl., vol. 7, serie 10, 287
(1958).}
\bibitem{MaNi}{Z. Q. Ma, G.-J. Ni, Phys. Rev. D {\bf31}, 1482 (1985).
See also Phys. Rev. D {\bf32}, 2203 (1985), Phys. Rev. D {\bf32},
2213 (1985).}
\bibitem{Barth}{M.-C. Barth\'el\'emy, Ann. Inst. Henri Poincar\'e,
{\bf7}, 115 (1967).}
\bibitem{Ni}{G.-J. Ni, Phys. Energ. Fortis {\&} Phys. Nucl. (China),
vol. 3, no. 4, p. 432-49 (1979).}
\bibitem{F-W}{J. D. Bjorken, S. D. Drell, {\it Relativistic Quantum
Mechanics} (McGraw-Hill, New York, 1964).}
\end{references}
\end{document}